\documentclass[conference]{IEEEtran}

\usepackage{graphicx}
\usepackage{amsmath}
\usepackage{amssymb}
\usepackage{enumitem}
\usepackage{hyperref}
\usepackage{microtype}
\usepackage{tikz}
\usepackage{xcolor}
\usepackage{booktabs}
\usepackage{multirow}
\usepackage{array}

\usepackage[htt]{hyphenat}

\usepackage{algorithm}
\usepackage{algorithmic}
\usepackage{float}

\usetikzlibrary{arrows.meta,positioning,fit,backgrounds,calc}

\title{
Dictionary-Guided Mutation Operators for Automated HDL Repair
}

\author{
\IEEEauthorblockN{Maisha Mastora}
\IEEEauthorblockA{
Department of Electrical and Computer Engineering\\
University of New Hampshire\\
Durham, NH, USA\\
maisha.mastora@unh.edu
}
\and
\IEEEauthorblockN{Dean Sullivan}
\IEEEauthorblockA{
Department of Electrical and Computer Engineering\\
University of New Hampshire\\
Durham, NH, USA\\
dean.sullivan@unh.edu
}
}

\IEEEoverridecommandlockouts
\IEEEpubid{\makebox[\columnwidth]{979-8-3195-3156-8/26/\$31.00~\copyright~2026 IEEE\hfill}\hspace{\columnsep}\makebox[\columnwidth]{}}

\begin{document}

\maketitle
\begin{center}\begin{tikzpicture}[remember picture, overlay]
\node[anchor=north] at ([yshift=-0.4cm]current page.north)
  {\small\textit{This paper has been accepted to 35th Microelectronics Design and test Symposium (MDTS) 2026}};
\end{tikzpicture}
\end{center}
\begin{abstract}
Automated repair of Hardware Description Language (HDL) designs remains challenging due to the large search space of candidate repairs and the strict syntactic and semantic constraints imposed by HDL grammars. Generic mutation strategies overwhelmingly generate syntactically invalid candidates that waste compilation and simulation budget, while synthesis-driven and template-based approaches impose their own constraints on generality and portability. In this paper, we propose a dictionary-guided HDL repair system that combines ANTLR-derived DUT-specific mutation vocabularies with a simulation-divergence fault localization (FL) module. The mutation operator applies category-constrained token substitutions, insertions, and deletions directly to Verilog source via regex-based matching, without requiring AST manipulation or synthesis. The FL module identifies diverging output wires from a single simulation run and scores source lines by structural proximity to those signals, directing the mutation search toward high-suspicion regions. A deterministic targeted sweep exhausts all dictionary mutations on the highest-scored lines before falling back to a genetic programming (GP) search. Evaluated on the CirFix benchmark suite across six design under test (DUT) families, the proposed approach produces correct oracle-passing repairs on 14 bug variants, including a 6-edit multi-bug instance that CirFix cannot repair, and achieves an 18$\times$ speedup over CirFix on a two-edit benchmark variant. These results indicate that dictionary-constrained mutation operators, combined with lightweight simulation-divergence FL, are a practical and competitive approach to automated HDL repair for common bug classes without formal analysis or synthesis dependencies.
\end{abstract}


\IEEEpubidadjcol
\section{Introduction}

Automated Program Repair (APR) has been studied extensively in the software domain, where techniques such as symbolic reasoning, constraint solving, and heuristic search have demonstrated the ability to produce correct patches with minimal human intervention~\cite{kim2023automated}. A parallel line of work treats repair as a search problem guided by an oracle, where the objective is to navigate a large space of program transformations toward variants that satisfy the required correctness specification. From this perspective, the quality of the search is determined not only by the oracle, but by the mutation operator that generates candidates for evaluation.

A fundamental limitation of text-level or byte-level mutation is poor alignment with program syntax. For languages with rich formal grammars — such as HDLs — unconstrained token edits produce syntactically invalid candidates at a high rate, exhausting compilation and simulation budget on variants that cannot contribute to repair. Prior work in structured fuzzing has demonstrated that token-aware mutation, which respects the syntactic structure of the input language, substantially improves exploration efficiency~\cite{mathis2020learning,salls2021token}. The same principle applies directly to HDL repair: constraining mutations to grammar-permitted, design-local tokens increases the density of valid and behaviorally plausible candidates in the search space.

HDLs are particularly well-suited to this approach. Their syntax and semantics are defined by formal grammars with explicit token classes, operator categories, and contextual constraints. These structures can be extracted statically from any design to construct a DUT-specific mutation vocabulary without simulation or synthesis. The vocabulary defines precisely what substitutions, insertions, and deletions are structurally meaningful for each token category in the design under repair.

In this paper, we present a complete HDL repair system built around this principle. The core contributions are: (i) an ANTLR-based dictionary construction pipeline that extracts DUT-specific structural and numeric token vocabularies from the Verilog grammar and the buggy source; (ii) a simulation-divergence FL module that scores source lines by proximity to diverging output wires without AST parsing; (iii) a category-constrained mutation operator that directs substitutions, insertions, and deletions toward high-suspicion regions; and (iv) a targeted deterministic sweep that exhausts single-edit repairs before engaging a GP fallback, enabling multi-edit repair through iterative partial-fix accumulation.

We evaluate our work on the CirFix benchmark suite~\cite{ahmad2022cirfix} across six DUT families, comparing against CirFix as a baseline. Our approach repairs 14 bug variants, including variants that CirFix cannot fix, and achieves substantially lower evaluation counts on several benchmarks. The results suggest that dictionary-constrained mutation, without formal analysis or synthesis, is a viable and competitive strategy for HDL repair across a range of common bug classes. Crucially, the mutation vocabulary is derived automatically from each design, requiring no manual repair templates.

\vspace{-1mm}
\section{Background and Related Work}

\subsection{HDL Repair and Oracle-Based Evaluation}
Automated repair of hardware design code differs fundamentally from software repair due to the concurrent, stateful, and timing-dependent nature of HDLs. Correctness is typically defined through simulation behavior against reference input-output traces, rather than sequential control flow properties. Consequently, HDL repair systems rely on explicit correctness oracles — testbenches or golden traces — to evaluate candidate fixes~\cite{ahmad2022cirfix,laeufer2024rtl}.

CirFix~\cite{ahmad2022cirfix} is the most directly comparable existing work. It applies a genetic programming search over AST-level mutations, guided by a weighted fitness function that scores candidate output traces against the oracle. CirFix operates on the full design AST, applying insertion, deletion, and replacement at the statement and expression level. While effective on several benchmark classes, CirFix incurs high per-evaluation cost due to AST manipulation, and its mutation operators are not constrained by the DUT's own token vocabulary. Certain fixes further require design-specific template declarations, limiting generalization across diverse designs.

RTL-Repair~\cite{laeufer2024rtl} takes a symbolic approach, encoding the repair problem as a bounded synthesis query over a localized fault window. This yields fast and correct repairs within its synthesis scope, but requires synthesizability of the target design and is limited to patches within the bounded template. SRepair~\cite{liu2025srepair} applies symbolic regression over signal traces to infer repair expressions, and STRIDER~\cite{yang2023strider} uses signal transition patterns to guide repair for specific defect classes. Both improve repair coverage for their target bug types but introduce synthesis or formal analysis dependencies that the proposed approach intentionally avoids.

\subsection{Fault Localization for Hardware}
Fault localization (FL) for HDL has been approached through both learning-based and simulation-based methods. Hu and Liu~\cite{hu2025context} use deep learning over execution coverage and contextual token representations to rank suspicious statements, achieving strong localization accuracy but leaving patch synthesis to a separate stage. STRIDER~\cite{yang2023strider} and related approaches integrate FL with repair by using signal-level divergence to constrain the search space.

The FL module proposed in this work is closest in spirit to simulation-divergence approaches: it identifies the first timestep at which output wires diverge from the oracle, then traces backward through the source using regex patterns to score lines by structural relationship to the diverging signal. This requires no AST, no learning, and only a single simulation run of the buggy design.

\subsection{LLM-Based Approaches}
Large language model-based approaches have recently been explored for hardware debugging~\cite{fu2024generalize}. These methods leverage pre-trained models to suggest patches from natural language or code context. While promising in coverage, they generally lack syntactic validity guarantees and are difficult to deploy in deterministic, oracle-constrained repair pipelines. The proposed system operates entirely within a verifiable, simulation-grounded framework.

\subsection{Structured Mutation in Software APR}
Token-level fuzzing~\cite{mathis2020learning,salls2021token} has demonstrated that mutation operators respecting input language structure substantially outperform byte-level approaches in generating valid, coverage-increasing inputs. Kim et al.~\cite{kim2023automated} formalize the connection between fuzzing-style search and APR, motivating oracle-guided mutation as a repair strategy. The proposed approach extends this line of reasoning to HDL repair, where the grammar structure is particularly rich and the cost of invalid candidates — compilation and simulation — is high.

\section{System Design}

\begin{figure*}[!t]
\centering
\begin{tikzpicture}[
  font=\small,
  every node/.style={align=center},
  topbox/.style={
    draw=blue!40, fill=blue!6, rounded corners=4pt,
    text width=2.6cm, minimum height=1.15cm,
    inner sep=5pt, font=\small\bfseries
  },
  repbox/.style={
    draw=teal!60, fill=teal!7, rounded corners=4pt,
    text width=2.6cm, minimum height=1.15cm,
    inner sep=5pt, font=\small\bfseries
  },
  sweepbox/.style={
    draw=teal!60, fill=teal!7, rounded corners=4pt,
    text width=3.0cm, minimum height=1.5cm,
    inner sep=5pt, font=\small\bfseries
  },
  graybox/.style={
    draw=gray!45, fill=gray!7, rounded corners=4pt,
    text width=2.2cm, minimum height=1.0cm,
    inner sep=5pt, font=\small
  },
  outbox/.style={
    draw=green!50!black, fill=green!7, rounded corners=4pt,
    text width=2.4cm, minimum height=1.15cm,
    inner sep=5pt, font=\small\bfseries
  },
  arr/.style={-latex, thick, draw=black!55},
  darr/.style={-latex, thick, draw=black!40, dashed},
]

\node[topbox] (grammar)
  {HDL grammar\\[1pt]{\footnotesize ANTLR / ATN}};

\node[topbox, right=9mm of grammar] (dut)
  {Buggy DUT\\[1pt]{\footnotesize Verilog source}};

\node[topbox, right=9mm of dut] (dictbuild)
  {Dictionary build\\[1pt]{\footnotesize STRUCT + NUM}};

\node[topbox, right=9mm of dictbuild] (tokdict)
  {Token dictionary\\[1pt]{\footnotesize DUT-specific vocab}};

\draw[arr] (grammar)   -- (dut);
\draw[arr] (dut)       -- (dictbuild);
\draw[arr] (dictbuild) -- (tokdict);

\begin{scope}[on background layer]
  \node[fill=blue!4, rounded corners=8pt, inner sep=9pt,
        fit=(grammar)(dut)(dictbuild)(tokdict)] (topgroup) {};
\end{scope}
\node[font=\footnotesize\bfseries, text=black!65,
      above=1pt of topgroup.north] {Dictionary construction (once per DUT)};

\node[graybox, below=20mm of grammar] (oracle)
  {Oracle\\[1pt]{\footnotesize golden trace}};

\node[repbox, right=9mm of oracle] (fl)
  {FL scoring\\[1pt]{\footnotesize single sim run}};

\node[sweepbox, right=9mm of fl] (sweep)
  {Targeted sweep\\[1pt]
   {\footnotesize all dict mutations\\
    on top-scored lines\\
    (up to $P\!=\!4$ passes)}};

\node[repbox, right=9mm of sweep] (gp)
  {GP fallback\\[1pt]{\footnotesize FL-biased mutation\\tournament select}};

\node[outbox, right=9mm of gp] (repaired)
  {Repaired DUT\\[1pt]{\footnotesize fitness $= 1.0$}};

\draw[arr] (oracle) -- (fl);
\draw[arr] (fl)     -- (sweep);
\draw[arr] (sweep)  -- node[above, font=\footnotesize\bfseries, text=black!65] {no fix} (gp);
\draw[arr] (gp)     -- (repaired);

\draw[arr] (sweep.north) -- ++(0, 0.7) -| (repaired.north);
\node[font=\footnotesize\bfseries, text=black!65]
  at ([xshift=4mm, yshift=4mm] sweep.north) [right] {fix found};

\draw[darr] (sweep.south) -- ++(0, -0.75)
    -| node[midway, below, font=\footnotesize\bfseries, text=black!65]
       {re-run FL on partial fix} (fl.south);

\draw[darr] (tokdict.south) -- ++(0, -5mm) -| (sweep.north);
\draw[darr] (tokdict.south) -- ++(0, -5mm) -| (gp.north);

\end{tikzpicture}
\caption{System overview. The dictionary construction stage (top, shaded) is executed once per DUT and produces a token vocabulary that guides all mutations. The repair pipeline (bottom) runs FL scoring from a single simulation, then exhaustively sweeps top-scored lines; if no fix is found the best partial candidate is passed to GP-based search.}
\label{fig:workflow}
\end{figure*}
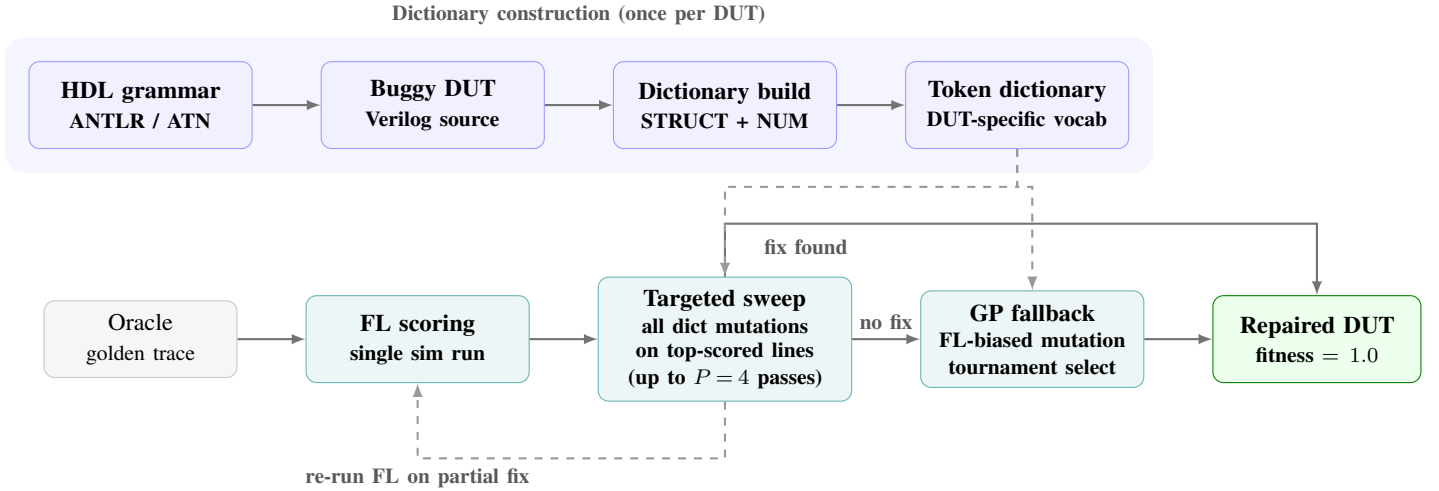

\subsection{Dictionary Construction}
\label{sec:dict}

As illustrated in Fig.~\ref{fig:workflow}, the repair pipeline begins by constructing a DUT-specific mutation vocabulary from the buggy Verilog source. Dictionary construction is performed once per DUT before any repair attempt, using ANTLR~\cite{antlr4github} to parse the design and analyze the grammar's Augmented Transition Network (ATN).

For each token in the buggy source, the ANTLR parse tree records its grammar rule context. We then traverse the ATN from each matched rule's start state to enumerate all literal tokens reachable within that syntactic context. This produces a \emph{context-aware} token vocabulary: rather than augmenting the design's tokens with the full set of Verilog keywords, we augment only with tokens that are grammatically valid replacements given the structural role of each source token. The result is two dictionaries per DUT: a \emph{structural dictionary} (STRUCT) containing identifiers, keywords, and operators, and a \emph{numeric dictionary} (NUM) containing valid Verilog constant literals.

Numeric literals are handled separately to preserve semantic validity. Binary literals in the source are used as seeds; for each seed, the NUM dict is populated with Hamming-distance-1 and distance-2 neighbors (single and double bit flips), all-zeros and all-ones variants, and one-position left and right rotations. Plain decimal integers (e.g., loop bounds) are augmented with a small neighborhood of adjacent values. No cross-width or cross-base substitutions are generated, preserving the original literal's type context.

When multiple buggy variants of the same DUT exist, per-variant dictionaries are merged to produce both a \emph{union} dictionary (covering all tokens seen across variants) and an \emph{intersection} dictionary (covering only tokens common to all variants). The union dictionary is used by default, maximizing vocabulary coverage.

\subsection{Simulation-Divergence Fault Localization}
\label{sec:fl}

We use simulation-divergence fault localization to identify suspicious lines. The buggy design is simulated once and its output CSV compared against the oracle to identify the first diverging timestep and the set of affected output wires. These wires anchor a backward trace through the source.

Analysis uses only regular expressions on raw Verilog text — no AST or parser. Each line receives a suspiciousness score in $[0.1, 1.0]$ based on its structural relationship to the diverging signals. Scoring proceeds in four passes. Pass~A scores any \texttt{assign} statement whose LHS includes a diverging wire at 1.0, implicating the full statement up to its terminating semicolon. Pass~B locates \texttt{always} blocks via \texttt{begin}/\texttt{end} depth counting. Blocks driving diverging wires score their header at 0.8 (or 1.0 if the sensitivity list lacks an edge keyword, indicating a missing \texttt{posedge}/\texttt{negedge} bug). Pass~C handles logic inside these blocks: \texttt{if}/\texttt{case} conditions score 0.7--0.9 and procedural assignments score 0.9 (reduced to 0.5 inside detected reset handlers, which rarely contain the functional bug). Pass~D first scores \texttt{always} blocks driving signals on the RHS of already-implicated lines ($\geq 0.7$) at 0.8 for headers and conditions, then assigns 0.3 to any line assigning to a secondary signal from high-scoring RHS expressions. All other lines default to 0.1.

The method is hardware-aware: it distinguishes combinational from sequential logic and treats sensitivity-list bugs as a first-class case by inspecting \texttt{always@} headers for edge qualifiers. If no divergence is detected, every line receives a uniform score of 0.5 and FL guidance is disabled.

\subsection{Mutation Operators}
\label{sec:mutops}

The mutation operator works directly on Verilog source text using pre-compiled regex patterns. Three primitives are defined:

\textbf{Substitute:} A token present in the source is replaced with a same-category token from the dictionary. Identifiers are further restricted to replacements that also appear in the source (preserving in-vocabulary semantics). Numeric tokens are drawn from the same-width pool in the NUM dict, with binary tokens sorted by Hamming distance so that single-bit-flip substitutions are sampled with higher probability. Keyword categories (e.g., \texttt{posedge}/\texttt{negedge}) are not source-filtered, since the absence of such a token \emph{is} the bug in sensitivity list defects. Substitution selects among numeric and structural candidates with equal probability (50/50 split) to prevent either category from dominating.

\textbf{Insert:} A keyword token is inserted immediately before or after a user identifier occurrence. The insertable pool is restricted to edge keywords (\texttt{posedge}/\texttt{negedge}), conditional trail tokens (\texttt{else}), and unary logical operators (\texttt{!}, \texttt{\textasciitilde}). When the insertion anchor is detected inside an \texttt{always@(...)} sensitivity list, the pool is further restricted to edge keywords only.

\textbf{Delete:} A token occurrence is removed. Only tokens in non-structural categories are eligible; identifiers, arithmetic/logical operators, block delimiters, module keywords, and net-type declarations are never deleted, as removal would leave syntactically broken or semantically undefined source. When FL line scores are available, all three operators prefer high-suspicion lines: each token in the source is assigned a score equal to the maximum line score across all its occurrences, and tokens are sorted descending by this score before the mutation loop begins. Among multiple occurrences of the selected token, the specific occurrence to mutate is sampled proportionally to the line suspiciousness scores, biasing mutations toward FL-implicated regions without hard-restricting the search.

\subsection{Repair Pipeline}
\label{sec:repair}

The repair procedure, summarized in Algorithm~\ref{alg:repair}, proceeds in two stages.

\begin{algorithm}[t]
\small
\caption{Repair Procedure}
\label{alg:repair}
\begin{algorithmic}[1]
\REQUIRE Buggy source $s_0$, mutation dictionary $\mathcal{D}$, oracle $\mathcal{O}$
\STATE $\text{scores} \leftarrow \text{FL}(s_0, \mathcal{O})$ \COMMENT{single buggy simulation}
\FOR{$p = 1$ \TO $P$}
  \STATE $s^* \leftarrow \text{TargetedSweep}(s_\text{cur}, \mathcal{D}, \text{scores})$
  \IF{$\text{fitness}(s^*) = 1.0$} \RETURN $s^*$ \ENDIF
  \IF{$\text{fitness}(s^*) > \text{fitness}(s_\text{cur})$}
    \STATE $s_\text{cur} \leftarrow s^*$; $\text{scores} \leftarrow \text{FL}(s_\text{cur}, \mathcal{O})$
  \ELSE \STATE \textbf{break}
  \ENDIF
\ENDFOR
\STATE $\text{pop} \leftarrow \text{SeedPopulation}(s_0, \mathcal{D}, \text{scores})$
\FOR{$g = 1$ \TO $G$}
  \STATE $\text{pop} \leftarrow \text{EvaluateAndSelect}(\text{pop}, \mathcal{D}, \text{scores})$
  \IF{any individual has fitness $1.0$} \RETURN it \ENDIF
\ENDFOR
\RETURN best individual seen
\end{algorithmic}
\end{algorithm}

\textbf{Stage 1: Targeted deterministic sweep.} Prior to stochastic search, the repair procedure exhaustively applies every dictionary mutation to the highest-scored lines. In each pass, every token on the top-scoring line is substituted with every same-category dictionary entry; every edge keyword is inserted before every identifier on that line; and every deletable conditional token is removed from all lines scoring $\geq 0.7$. The sweep also covers all lines scoring $\geq 0.7$ with full substitution, applies unary operator insertion (\texttt{!}, \texttt{\textasciitilde}) across top and mid-score lines, and performs identifier and numeric substitution on lines scoring in $[0.3, 0.6)$. Each candidate is evaluated immediately and the sweep exits upon the first fitness-1.0 result.

When no candidate achieves fitness 1.0 but one strictly improves over the current best, FL is re-executed on that improved partial candidate. The updated scores reflect the residual divergence after the partial fix, potentially implicating different lines than the original scoring. The sweep is then re-executed on the updated candidate with the new scores, excluding previously swept lines to avoid redundant work. This repeats for up to $P$ passes (default $P = 4$). Each pass operates on the best candidate produced by the preceding pass, allowing independently discovered single-edit fixes to accumulate. This yields effective multi-edit repair without requiring coordinated multi-token mutations within a single individual — a known difficulty in standard GP-based APR~\cite{ahmad2022cirfix}. If all passes are exhausted without a perfect repair, the best partial candidate is retained and control passes to Stage~2.

\textbf{Stage 2: GP fallback.} A population of \texttt{POPSIZE} candidates (default 200) is seeded by applying FL-biased single mutations to the original buggy source. Seeding from the original source maintains vocabulary diversity and prevents premature convergence on partial fixes. Each generation applies tournament selection (tournament size $k=3$) to select parents, then applies a single FL-biased mutation at rate \texttt{MUTATION\_RATE} (default 0.7). The top 10\% of each generation are carried forward unchanged (elitism). Any candidate achieving fitness 1.0 terminates the search immediately. The search runs for \texttt{GENS} generations (default 10) across \texttt{RESTARTS} independent restarts (default 5).

\textbf{Fitness function.} Fitness is computed by the CirFix oracle fitness function~\cite{ahmad2022cirfix}: each candidate is compiled with \texttt{iverilog}, simulated, and its output CSV compared bitwise against the oracle trace. Matching bits contribute $+1$, mismatches $-1$, and \texttt{x}/\texttt{z} matches contribute $\pm 2$ (weighted). The normalized fitness $f \in [0, 1]$ is returned; candidates that fail compilation receive fitness 0 without simulation.

\vspace{-1mm}
\section{Experimental Evaluation}

\subsection{Setup}
\label{sec:setup}

We evaluate the proposed system on the CirFix benchmark suite~\cite{ahmad2022cirfix}, which provides buggy Verilog designs with corresponding testbenches and golden oracle traces. We use the following six DUT families: \texttt{first\_counter\_overflow}, \texttt{fsm\_full}, \texttt{lshift\_reg}, \texttt{mux\_4\_1}, \texttt{decoder\_3\_to\_8}, and \texttt{flip\_flop}. Designs where all variants pass the provided oracle with fitness 1.0 on the original buggy source are excluded, as this indicates insufficient testbench coverage to expose the injected bugs — a known oracle weakness in the benchmark, not a failure of the repair system.

The repair system is configured with \texttt{popsize=200}, \texttt{gens=10}, \texttt{restarts=5}, \texttt{mutation\_rate=0.7}, and \texttt{dict\_type=union}. To ensure a fair comparison under identical conditions, we re-ran the original CirFix implementation on the same machine and simulator used for all experiments. The original CirFix paper used Synopsys VCS, a commercial simulator significantly faster per simulation cycle than \texttt{iverilog}. Because we do not have access to VCS, all local runs use \texttt{iverilog} for both systems, making wall-clock times directly comparable on our setup. Paper-reported CirFix times are cited where local replication was not feasible, but these are not directly comparable due to the simulator difference and are marked accordingly.

We use \emph{fitness evaluation count} as the primary efficiency metric, since it is machine- and simulator-independent: one evaluation corresponds to a complete compile-simulate-score cycle. Wall-clock times on the same machine and simulator are reported as a secondary metric where available.

All experiments were conducted on a Lenovo ThinkPad P16 Gen~1 (x86-64, Ubuntu 25.10) with Python~3.9 and \texttt{iverilog} as the simulator. For non-template bugs requiring large-scale GP search, vanilla CirFix with \texttt{popsize=5000} did not complete within a practical time limit under \texttt{iverilog}. This reflects a fundamental difference in search strategy: CirFix requires a large population to compensate for undirected mutation, whereas the proposed system repairs the same bugs with \texttt{popsize=200} because dictionary-guided mutation restricts candidates to semantically plausible token substitutions from the design's own vocabulary.

\subsection{Results}
\label{sec:results}

Table~\ref{tab:results} summarizes repair outcomes across all evaluated variants. Out-of-scope variants — those requiring repairs outside the current mutation vocabulary (e.g., commented-out code, hex-to-binary base conversions, port declaration changes) — are listed separately and excluded from the repair count.

\begin{table*}[t]
\centering
\small
\caption{Repair results on the CirFix benchmark suite.
OOS = out of scope. $\dagger$ = oracle-passing; weak oracle (Section~\ref{sec:discussion}).
$^\text{u}$ = unique fix not repaired by CirFix.
Times in seconds; CirFix times are local \texttt{iverilog} re-runs where available, otherwise
paper-reported VCS times$^*$ (not directly comparable). --- = not attempted or did not finish.}
\label{tab:results}
\renewcommand{\arraystretch}{1.03}
\setlength{\tabcolsep}{4pt}
\begin{tabular}{@{}llp{5.0cm}cc@{}}
\toprule
\textbf{DUT} & \textbf{Variant} & \textbf{Bug Type} & \textbf{Result} & \textbf{CirFix} \\
\midrule
\multirow{6}{*}{\textit{first\_counter}}
 & buggy\_overflow  & Numeric sub: \texttt{1'b0}$\to$\texttt{1'b1}                              & \checkmark\ 0.56s              & \checkmark\ 0.4s$^*$ \\
 & wadden\_buggy1   & Missing \texttt{posedge} in sensitivity list                               & \checkmark\ 2.04s              & \checkmark\ 2.64s \\
 & wadden\_buggy2   & Spurious \texttt{else} branch                                              & \checkmark\ 7.93s              & \checkmark\ 2.84s \\
 & buggy\_counter   & 3-edit: \texttt{+2}$\to$\texttt{+1}, \texttt{>=}$\to$\texttt{==}, bit fix  & \checkmark\ 20.9s$^\text{u}$   & $\times$ \\
 & kgoliya\_buggy1  & Commented-out reset line                                                   & OOS                            & --- \\
 & buggy\_all       & 3 simultaneous edits (combined)                                            & OOS                            & --- \\
\midrule
\multirow{5}{*}{\textit{lshift\_reg}}
 & kgoliya\_buggy1  & Edge keyword: \texttt{negedge}$\to$\texttt{posedge}                        & \checkmark\ 38.7s            & \checkmark\ 7.8s \\
 & buggy\_num       & Plain int: \texttt{i=2}$\to$\texttt{i=0}                                   & \checkmark\ 13.3s$^\text{u}$ & --- \\
 & buggy\_var       & Identifier sub: \texttt{rstn}$\to$\texttt{load\_en}                        & \checkmark\ 25.1s            & \checkmark\ 33.7s \\
 & wadden\_buggy2   & Missing \texttt{!} before \texttt{rstn}                                    & \checkmark\ 181s$^\text{u}$  & --- \\
 & wadden\_buggy1   & 4 blocking \texttt{=}$\to$\texttt{<=}                                      & $\times$ (best 0.83)         & \checkmark\ 14.6s \\
\midrule
\multirow{4}{*}{\textit{mux\_4\_1}}
 & buggy\_var       & Identifier sub: \texttt{a}$\to$\texttt{b}                                 & \checkmark\ 14.5s$^\text{u}$ & $\times$ \\
 & wadden\_buggy1   & 3 case labels wrong (simultaneous)                                         & $\times$ (best 0.73)         & $\times$ \\
 & kgoliya\_buggy1  & Missing \texttt{[3:0]} on output port                                      & OOS                          & --- \\
 & wadden\_buggy2   & Hex$\to$binary conversion (4 edits)                                        & OOS                          & --- \\
\midrule
\multirow{6}{*}{\textit{decoder\_3\_to\_8}}
 & buggy\_var       & Identifier sub: \texttt{en}$\to$\texttt{C}                                & \checkmark\ 34.1s$^\text{u}$ & $\times$ \\
 & wadden\_buggy1   & 2-edit numeric                                                             & \checkmark\ \textbf{776s}    & \checkmark\ 13984s$^*$ \\
 & buggy\_num       & Numeric sub: \texttt{8'b1111\_1111}$\to$\texttt{...1110}                  & \checkmark\ 484s$^\text{u}$  & --- \\
 & super\_buggy     & 6-edit multi-bug (4-pass sweep)                                            & \checkmark\ 2110s$^\text{u}$ & $\times$ \\
 & wadden\_buggy2   & Missing underscores in literals                                            & OOS                          & --- \\
 & kgoliya\_buggy1  & Syntax error: \texttt{assign <=}                                           & OOS                          & --- \\
\midrule
\multirow{5}{*}{\textit{fsm\_full}}
 & buggy\_num       & Numeric sub: \texttt{1'b1}$\to$\texttt{1'b0}                              & \checkmark$\dagger$          & \checkmark \\
 & buggy\_var       & Identifier sub: \texttt{GNT1}$\to$\texttt{IDLE}                           & \checkmark$\dagger$          & \checkmark \\
 & ssscrazy\_buggy1 & Missing \texttt{state} in sensitivity list                                 & \checkmark$\dagger$          & \checkmark \\
 & ssscrazy\_buggy2 & Missing assignment + default case                                          & $\times$ (best 0.99)         & $\times$ \\
 & wadden\_buggy1   & Commented-out case branch                                                  & OOS                          & --- \\
\midrule
\multirow{2}{*}{\textit{flip\_flop}}
 & tff\_wadden\_buggy1 & Missing \texttt{!} before \texttt{rstn}                                & \checkmark\ 8.78s            & \checkmark\ 7.8s \\
 & tff\_wadden\_buggy2 & Swapped branches (simultaneous 2-edit)                                 & $\times$ (best 0.78)         & \checkmark\ 924s$^*$ \\
\bottomrule
\end{tabular}
\end{table*}

\paragraph{Overall repair count}
The proposed system produces oracle-passing repairs on \textbf{14 bug variants} across five DUT families, from a total of 23 in-scope variants (61\% repair rate). Three additional variants fail due to limitations discussed in Section~\ref{sec:discussion}, and six are out of scope for the current mutation vocabulary.

\paragraph{Multi-edit repair}
The \texttt{decoder\_3\_to\_8} super\_buggy variant contains six simultaneous bugs. The proposed system repairs it in four targeted sweep passes (5374 total evaluations), with each pass accumulating one or more single-edit fixes on the best partial candidate from the preceding pass. CirFix does not repair this variant. The \texttt{first\_counter} buggy\_counter variant (3 edits) is similarly repaired via the iterative targeted sweep in 98 evaluations.

\paragraph{Efficiency and unique fixes}
On the \texttt{decoder\_3\_to\_8} wadden\_buggy1 variant, the proposed system requires 776s locally versus the paper-reported 13{,}984s for CirFix (obtained on VCS) — an \textbf{18$\times$ speedup}. This comparison is noted with the caveat that the CirFix figure uses a faster commercial simulator; the evaluation count comparison (1951 vs.\ the thousands required by CirFix's GP search) is simulator-independent and reflects the same advantage. For non-template bugs, local vanilla CirFix with \texttt{popsize=5000} did not finish under \texttt{iverilog} within a practical time limit, while the proposed system resolves these with \texttt{popsize=200}. Five additional variants marked $^\text{u}$ are repaired by the proposed system but not by CirFix, covering identifier substitution, plain integer, and unary operator bug classes outside CirFix's mutation template coverage. Our system trades some speed on template-covered bugs for generality: its mutation vocabulary is derived automatically from any design, requiring no manually curated templates.
\vspace{-1mm}
\section{Discussion and Limitations}
\label{sec:discussion}

\paragraph{Simulator and timing comparability}
The original CirFix paper used Synopsys VCS; our evaluation uses \texttt{iverilog}, which has higher per-invocation overhead. All local comparisons between the two systems use \texttt{iverilog} on the same machine and are internally consistent. Paper-reported CirFix times are cited only where local re-runs were not feasible and are marked $^*$ accordingly; these are not directly comparable in wall-clock terms.

\paragraph{Oracle quality and spurious repairs}
Three \texttt{fsm\_full} variants are marked as oracle-passing with a weak oracle caveat. The \texttt{fsm\_full} testbench covers only 39 input combinations, which is insufficient to distinguish semantically correct repairs from spurious oracle-passing patches. The proposed system found oracle-passing repairs for \texttt{buggy\_num} and \texttt{buggy\_var} that differ from the intended fix (e.g., \texttt{req\_0}$\to$\texttt{gnt\_0} instead of the correct \texttt{GNT1}$\to$\texttt{IDLE}). The semantically correct fixes also pass the oracle, confirming that the oracle weakness — not the repair search — is the source of ambiguity. This is the standard \emph{overfitting} problem in oracle-guided APR, well-documented for both software~\cite{kim2023automated} and hardware repair systems~\cite{ahmad2022cirfix}. The proposed approach produces oracle-passing repairs; semantic correctness beyond the provided test suite is outside the scope of the repair guarantee.

\paragraph{Simultaneous multi-edit repair}
The \texttt{flip\_flop} tff\_wadden\_buggy2 variant requires a simultaneous swap of two assignments (\texttt{q<=q} $\leftrightarrow$ \texttt{q<={\textasciitilde}q}). The FL module correctly scores both buggy lines at 0.9, but the greedy single-edit targeted sweep fails: repairing either line in isolation reduces fitness before the other is corrected. CirFix handles this via AST-level GP, which tolerates fitness valleys. Similarly, the \texttt{mux\_4\_1} wadden\_buggy1 variant requires correcting three \texttt{case} labels simultaneously; any single-label fix temporarily reduces fitness below the buggy baseline (0.2667 vs.\ 0.7333). Both cases represent a fundamental limitation of greedy sequential repair and motivate coordinated multi-token patch generation as a direction for future work.

\paragraph{Blocking-to-non-blocking conversion}
The \texttt{lshift\_reg} wadden\_buggy1 variant requires converting four blocking assignments (\texttt{=}) to non-blocking (\texttt{<=}). The proposed system achieves a best fitness of 0.8276 — one assignment is correctly converted by the targeted sweep, but the GP search fails to accumulate all four fixes simultaneously, as the repair requires modifying four independent lines at once.

\paragraph{Non-minimal patches}
In the \texttt{lshift\_reg} kgoliya\_buggy1 variant, the correct single-edit fix (\texttt{negedge}$\to$\texttt{posedge}) is found in the first targeted sweep pass. However, subsequent passes introduce a spurious additional substitution (\texttt{i<8}$\to$\texttt{i<10}) that also passes the oracle. The resulting patch is functionally correct but not minimal. The current system does not include a post-processing minimization step; delta debugging~\cite{kim2023automated} over the produced patch would address this.

\paragraph{Out-of-scope mutations}
Several benchmark variants require repair operations outside the current vocabulary: commented-out code (requiring line reinsertion), port declaration width changes, hex-to-binary base conversions for literal constants not in the NUM dict, and compile-time syntax errors that prevent oracle evaluation. Extending the dictionary construction to cover these classes represents a clear path to broader coverage.

\vspace{-1mm}
\section{Conclusion and Future Work}

We presented a dictionary-guided HDL repair system combining ANTLR-derived DUT-specific mutation vocabularies with simulation-divergence fault localization. The system operates directly on Verilog source text without AST manipulation, synthesis, or formal analysis, making it portable across designs and independent of synthesizability constraints. The dictionary pipeline, executed once per DUT, bounds the mutation space to grammar-valid, design-specific tokens. The FL module, requiring only a single simulation of the buggy design, scores source lines by structural proximity to diverging output wires and directs both the targeted sweep and the GP fallback toward high-suspicion regions.

Evaluated on the CirFix benchmark suite across six DUT families, the proposed system produces oracle-passing repairs on 14 bug variants — including 3-edit and 6-edit multi-bug instances that CirFix cannot repair — while achieving up to an 18$\times$ speedup on shared benchmarks. These results indicate that dictionary-constrained mutation directed by lightweight simulation-divergence FL is a practical and competitive approach to HDL repair for common bug classes, without requiring formal analysis or synthesis.

Future work includes coordinated multi-token patch operators for simultaneous multi-edit bugs, post-repair minimization via delta debugging to eliminate spurious edits, extended numeric dictionary coverage for base-agnostic and signed arithmetic substitutions, and weaker oracle dependence through partial or property-based specifications.

\vspace{-1mm}
\bibliographystyle{IEEEtran}
\bibliography{references}

@inproceedings{kim2023automated,
  title={Automated program repair from fuzzing perspective},
  author={Kim, YoungJae and Han, Seungheon and Khamit, Askar Yeltayuly and Yi, Jooyong},
  booktitle={Proceedings of the 32nd ACM SIGSOFT International Symposium on Software Testing and Analysis},
  pages={854--866},
  year={2023}
}

@inproceedings{mathis2020learning,
  title={Learning input tokens for effective fuzzing},
  author={Mathis, Bj{\"o}rn and Gopinath, Rahul and Zeller, Andreas},
  booktitle={Proceedings of the 29th ACM SIGSOFT international symposium on software testing and analysis},
  pages={27--37},
  year={2020}
}

@inproceedings{salls2021token,
  title={$\{$Token-Level$\}$ Fuzzing},
  author={Salls, Christopher and Jindal, Chani and Corina, Jake and Kruegel, Christopher and Vigna, Giovanni},
  booktitle={30th USENIX Security Symposium (USENIX Security 21)},
  pages={2795--2809},
  year={2021}
}

@inproceedings{ahmad2022cirfix,
  title={CirFix: Automatically repairing defects in hardware design code},
  author={Ahmad, Hammad and Huang, Yu and Weimer, Westley},
  booktitle={Proceedings of the 27th ACM International Conference on Architectural Support for Programming Languages and Operating Systems},
  pages={990--1003},
  year={2022}
}

@inproceedings{laeufer2024rtl,
  title={Rtl-repair: Fast symbolic repair of hardware design code},
  author={Laeufer, Kevin and Fajardo, Brandon and Ahuja, Abhik and Iyer, Vighnesh and Nikoli{\'c}, Borivoje and Sen, Koushik},
  booktitle={Proceedings of the 29th ACM International Conference on Architectural Support for Programming Languages and Operating Systems, Volume 3},
  pages={867--881},
  year={2024}
}

@article{liu2025srepair,
  title={SRepair: Symbolic Regression-Based Repair for Hardware Design Code},
  author={Liu, Zizhen and Yang, Deheng and Mao, Xiaoguang and He, Jiayu and Zhang, Guangda and Lei, Yan and Wu, Jiang},
  journal={IEEE Transactions on Computer-Aided Design of Integrated Circuits and Systems},
  year={2025},
  publisher={IEEE}
}

@article{yang2023strider,
  title={STRIDER: Signal value transition-guided defect repair for HDL programming assignments},
  author={Yang, Deheng and He, Jiayu and Mao, Xiaoguang and Li, Tun and Lei, Yan and Yi, Xin and Wu, Jiang},
  journal={IEEE Transactions on Computer-Aided Design of Integrated Circuits and Systems},
  volume={43},
  number={5},
  pages={1594--1607},
  year={2023},
  publisher={IEEE}
}

@article{fu2024generalize,
  title={A generalize hardware debugging approach for large language models semi-synthetic, datasets},
  author={Fu, Weimin and Li, Shijie and Zhao, Yifang and Yang, Kaichen and Zhang, Xuan and Jin, Yier and Guo, Xiaolong},
  journal={IEEE Transactions on Circuits and Systems I: Regular Papers},
  year={2024},
  publisher={IEEE}
}

@article{hu2025context,
  title={Context Aware Deep Learning-Based Fault Localization for Hardware Design Code},
  author={Hu, Jian and Liu, Zhenlei},
  journal={IEEE Transactions on Computer-Aided Design of Integrated Circuits and Systems},
  year={2025},
  publisher={IEEE}
}

@misc{antlr4github,
  author       = {{ANTLR Project}},
  title        = {ANTLR4},
  howpublished = {\url{https://github.com/antlr/antlr4}},
  year         = {2024},
  note         = {Accessed: 2026-01}
}

\end{document}